\documentclass{emulateapj}

\def\xmm{{XMM-{\it Newton\/}}}
\def\fermi{{{\it Fermi}-LAT}}

\newcommand{\fhl}{2FHL~J0826.1$-$4500}

\shorttitle{Discovery of shock-cloud interaction on Western edge of Vela SNR}

\shortauthors{Eagle et al.}

\PassOptionsToPackage{colorlinks,linkcolor={blue},citecolor={blue},urlcolor={red}}{hyperref}
\usepackage{hyperref}
\usepackage{natbib}
\usepackage{float}
\usepackage{color}
\usepackage{graphicx}
\usepackage{gensymb}
\usepackage{enumitem}
\usepackage{multirow,helvet}

\begin{document}



\title{2FHL~J0826.1$-$4500: Discovery of a possible shock-cloud interaction on the Western edge of the Vela SNR}

\author{J. Eagle\altaffilmark{1}, S. Marchesi\altaffilmark{1}, D. Castro\altaffilmark{2}, M. Ajello\altaffilmark{1}, L. Duvidovich\altaffilmark{3}$^{,}$\altaffilmark{4}, L. Tibaldo\altaffilmark{5}}

\altaffiltext{1}{Department of Physics \& Astronomy, Clemson University, Clemson, SC 29634, USA}
\altaffiltext{2}{Harvard-Smithsonian Center for Astrophysics, Cambridge, MA 02138, USA}
\altaffiltext{3}{Universidad de Buenos Aires,Facultad de Cs Exactas y Naturales, Buenos Aires, Argentina}
\altaffiltext{4}{CONICET - Universidad de Buenos Aires, Instituto de Astronomía y Física del Espacio (IAFE), Buenos Aires, Argentina}
\altaffiltext{5}{IRAP/Observatoire Midi-Pyrénées, France}


\begin{abstract}
We report on the investigation of a very high energy (VHE), Galactic $\gamma$-ray source recently discovered at $>$50\,GeV using the Large Area Telescope (LAT) on board \textit{Fermi}. This object, 
2FHL~J0826.1$-$4500, displays one of the hardest $>$50\,GeV spectra (photon index $\Gamma_{\gamma}\sim1.6$) in the 2FHL catalog, and a follow-up observation with XMM-\textit{Newton} has uncovered diffuse, soft thermal emission at the position of the $\gamma$-ray source. 
A detailed analysis of the available multi-wavelength data shows that this source is located on the Western edge of the Vela supernova remnant (SNR): the observations and the spectral energy distribution modeling support a scenario where this $\gamma$-ray source is the byproduct of the interaction between the SNR shock and a neutral Hydrogen cloud. If confirmed, this shock-cloud interaction would make 2FHL~J0826.1$-$4500 a promising candidate for efficient particle acceleration.
\end{abstract}

\section{Introduction}
The plane of the Milky Way is rich with efficient accelerators of cosmic rays (CRs) whose interaction with the ambient medium and photon fields produces energetic $\gamma$-rays. As such, $\gamma$-rays represent an excellent probe of non-thermal astrophysical processes.  Relativistic electrons can produce $\gamma$-rays by non-thermal bremsstrahlung or by inverse Compton scattering (IC) on ambient photon fields, whereas protons and heavier nuclei (i.e., hadrons) can generate $\gamma$-rays by the process of pion decay produced in collisions between relativistic hadrons and ambient material. Studies of the non-thermal Galactic source population are essential to understand how and where the bulk of cosmic rays are being accelerated and to understand the mechanisms underlying very high energy (VHE, E$>$50\,GeV) emitters \citep{renaud2009,karg2013}.

Several deep observations have been performed to study the Galactic plane in the TeV $\gamma$-ray energy band with facilities like the H.E.S.S., MAGIC, and VERITAS ground-based Cerenkov telescopes \citep{hess2003,magic2005,ver2006,antonelli2009}. These surveys led to the discovery that the Galactic plane is rich with TeV $\gamma$-ray emission from objects leftover after supernova explosions, such as pulsar wind nebulae (PWNe) and supernova remnants \citep[SNRs,][]{funk2005, aha2006, carrigan2013, ong2014}.

Recently, the Pass 8 \citep{atwood2013} event level reconstruction and analysis has enabled the \textit{Fermi}-Large Area Telescope (LAT) to achieve comparable performances to the aforementioned facilities at energies above 50\,GeV, reaching an average sensitivity in the plane of $\sim2\%$ of the Crab flux \citep[only slightly less sensitive than H.E.S.S. in this energy energy band, see][]{hess2018} with a localization accuracy better than 3$^\prime$ for most sources \citep{atwoodfermi}. \textit{Fermi}'s main advantage is that it has surveyed the entire sky, and hence the Galactic plane, with uniform sensitivity and coverage whereas other telescopes are limited to detection from ground-based locations (e.g. H.E.S.S., VERITAS, and MAGIC are all ground-based) and are restricted from viewing the entire Galactic plane with uniform sensitivity. As a result, the \textit{Fermi}-LAT \citep{atwood2013, pass82017} has detected several new Galactic sources, some of which display very hard spectra above 50\,GeV, which is a sign of efficient particle acceleration and (or) effective particle and energy dissipation processes. Understanding the properties of the VHE Galactic source population is crucial in order to identify the locations and mechanisms for Galactic cosmic ray acceleration. 

One breakthrough that Pass 8 has enabled has been the census of the entire sky at $>$50\,GeV reported in the 2FHL catalog \citep[][]{ackermann2015}, which is comprised of 360 sources detected across the entire sky. Of these objects, 103  are detected in the Galactic plane ($\mid b\mid$ $< 10\degree$): 38 of these have been associated with Galactic objects as their counterparts, 42 are associated with blazars, and 23 are unassociated. 

While none of these 23 unassociated sources has the radio and optical properties of blazars, it might still be possible to find $\gamma$-ray blazars on the Galactic plane that are undetected above the threshold of current radio surveys. A further selection criterion to classify sources as Galactic in origin is the hardness of the $\gamma$-ray spectrum at $\>$50\,GeV, since at these energies blazars generally exhibit  a soft spectrum (average photon index $\Gamma\sim3.4$), because the energy range is above the IC peak of their spectral energy distribution (SED). This is a result of the combination of the spectral shape of the energy distribution of the accelerated particles and the absorption due to the extragalactic background light \citep{dominguez2011}, which results in an exponentially cut-off photon spectrum. Only $\sim$4\,\% of the 2FHL blazars display a power law photon index $\Gamma<1.8$.
Among the 23 unidentified 2FHL objects located in the Galactic plane, 11 have $\Gamma<1.8$, and hence the number of contaminant blazars in this hard-spectrum sub-sample is expected to be $<1$. This sub-sample should therefore be mostly comprised of newly detected hard-spectrum Galactic objects. In this work, we focus on one of these 11 sources, 2FHL~J0826.1-4500, which is located at $\sim$1.5\degree\ South-West of the Vela pulsar, PSR J0835--4510.

Vela is among the closest SNRs to Earth, being at a distance d$\approx$290\,pc \citep{dodson_2003}, and it houses a middle-aged pulsar (characteristic age $t\approx11$\,kyr). Given its significant complexity, the Vela region has been widely studied in the literature.
The Vela pulsar sits in the central region of the SNR shell while actively fueling a large pulsar wind nebula (PWN) $2\degree \times 3\degree$ in size known as Vela-X. Generally, the structure of a composite SNR is complex and heavily depends on the density of the surrounding material it expands into. 

\citet{pslane} discusses how the layers of an SNR expand through a circumstellar medium (CSM) with a density gradient. The expansion of the PWN into the SNR, and the SNR expansion into the ISM are both responsible for heating ejecta and ambient material. The ejecta can confine the PWN, and as the PWN expands into the ejecta it also drives a shock, heating the material and producing thermal emission. 
The outer boundary of the SNR is defined by a forward shock (FS), which is the result of the material ejected from the initial explosion sweeping up the surrounding medium. It has been established that most massive stars often collapse and explode at a relatively early age, and thus they are believed to leave behind supernova remnants inside or nearby the dense molecular gas clouds where they were formed. It is not uncommon then, for these SNRs to interact with this dense gas regions as the shock expands. The morphology and spectral features of the forward shock can provide information regarding such an interaction.

This paper describes the analysis of newly acquired X-ray observations, as well as archival multi-wavelength data in the region where 2FHL~J0826.1$-$4500 is located and the modeling of the broadband spectral energy distribution to better understand the origin of the $\gamma$-ray emission. This paper is organized as follows: in Section \ref{sec:data} we discuss the source selection and the XMM-\textit{Newton} data reduction and analysis. A further, multi-wavelength characterization of the source is presented in Section \ref{sec:multi}. Section \ref{sec:discuss} explores the shock-cloud scenario through the SED modeling, and Section \ref{sec:conclude} summarizes our results.

\section{Source selection and data reduction}\label{sec:data}
\begin{figure*}
\begin{minipage}[b]{.5\textwidth}
  \centering
  \includegraphics[width=0.95\textwidth]{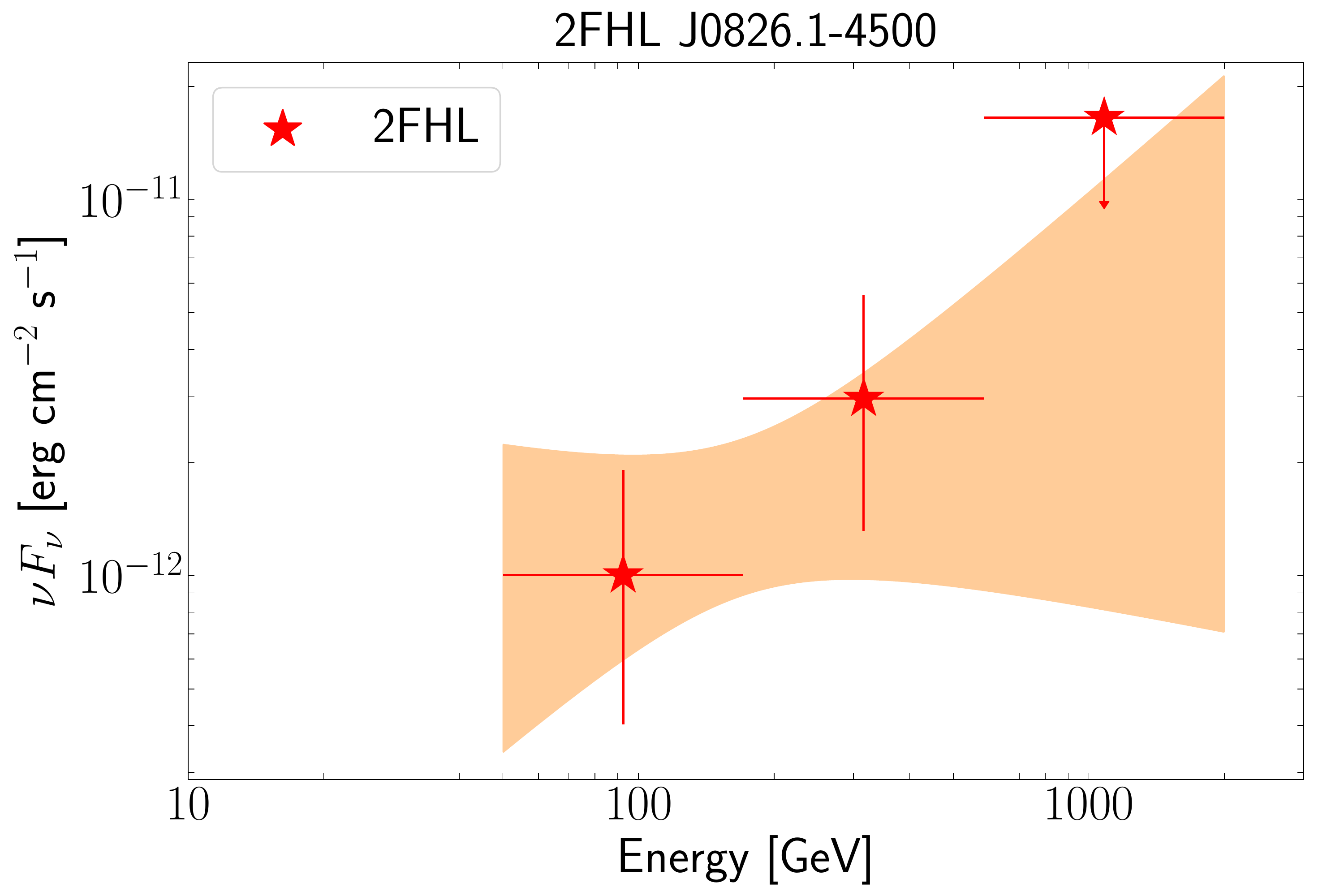}
  \end{minipage}
    \vspace{-0.5cm}
\begin{minipage}[b]{.5\textwidth}
  \centering
  \includegraphics[width=0.87\textwidth]{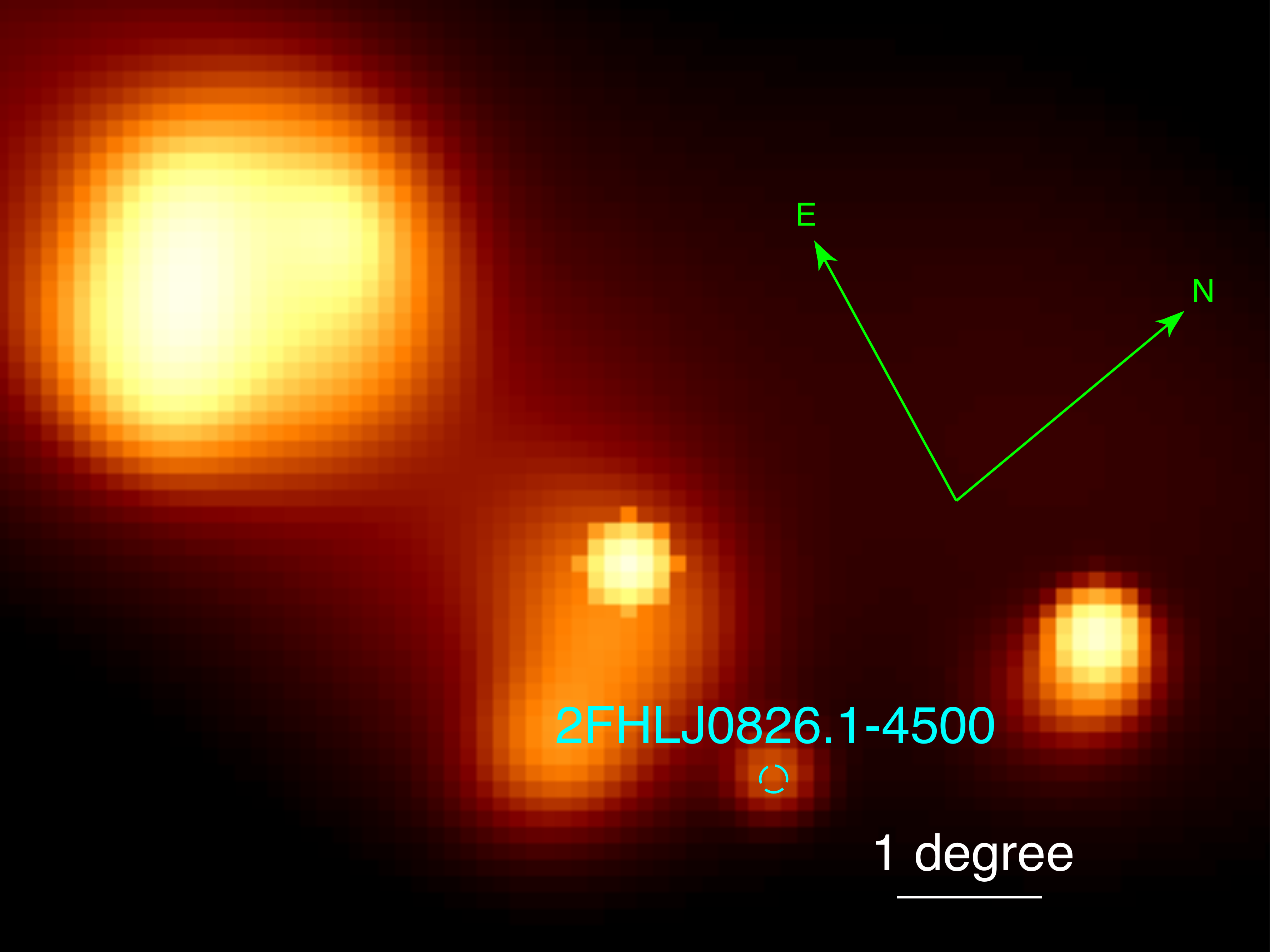}
  \end{minipage}
  \vspace{0.1cm}
\caption{\textit{Left}:  $\gamma$-ray SED of  2FHL~J0826.1$-$4500, using data from the 2FHL catalog \citep{ackermann2015}.
\textit{Right}: $\gamma$-ray image of the Vela complex at $\geq$50\,GeV. 2FHL~J0826.1$-$4500 shows no significant evidence of extended emission.}\label{fig:sed_gamma_map}
\end{figure*}

2FHL~J0826.1$-$4500 was first detected at $>$50\,GeV in the 2FHL catalog and presents a particularly hard $\gamma$-ray spectrum with photon index $\Gamma_\gamma=1.6\pm0.3$ and a maximum energy photon  of $\sim$412\,GeV as reported in the 2FHL catalog \citep{ackermann2015}. We report the source $>$50\,GeV spectrum in Figure \ref{fig:sed_gamma_map}, left panel. The source is compact and shows no clear evidence of extended emission beyond the point spread function of the \textit{Fermi}-LAT in this energy range (Figure \ref{fig:sed_gamma_map}, right panel).  

To further investigate the properties of this intriguing VHE object, we were granted a 20\,ks \xmm\ follow-up observation (proposal ID: 0782170201, PI: M. Ajello). X-ray telescopes like XMM-\textit{Newton} play a pivotal role in identifying Galactic $\gamma$-ray sources, as they provide arcsecond-scale angular resolution allowing the identification of the correct counterparts in the crowded Galactic region \citep{lumb2012}. Furthermore, XMM-\textit{Newton} also has the largest effective area in the 0.5--10\,keV band among all the X-ray telescopes, therefore being the most effective instrument to detect faint, diffuse X-ray emission along the Galactic plane, like the one commonly observed in PWNe and SNRs.
We report a summary of the observation details in Table \ref{tab:obsdetails}.

\begingroup
\renewcommand*{\arraystretch}{1.2}
\begin{table*}
\centering
\scalebox{1.}{
\begin{tabular}{ccccccccc}
\hline
\hline
\ Name & R.A. & Dec. & Obs. Date & Exp.$^a$ & Target Type & $\Gamma^b$ & $\Delta r^c$ & S/N$^d$ \
\\
\hline
2FHL~J0826.1$-$4500 & 08h 25m 56.63s & -45d 00$^\prime$ 00.0$^{\prime\prime}$ & 11/23/16 & 18500 & SNR & 1.6 $\pm0.3$ & 4.0 & 5.2 \\
\hline
\hline 
\end{tabular}}
\caption{Observation details of 2FHL~J0826.1$-$4500. {\footnotesize \textit{$^{a}$ Exposure time in $s$, $^{b}$ Photon index at $E>$50\,GeV, $^{c}$ Positional uncertainty of the $\gamma$-ray source (95$\%$ C.L.) in arc-minutes, $^{d}$ Signal to noise ratio of the $\gamma$-ray source ($\sigma$)}}}\label{tab:obsdetails}
\end{table*}
\endgroup

\subsection{\xmm\ Data Reduction and Analysis}
\begin{figure*}
\begin{minipage}[b]{.5\textwidth}
  \centering
\includegraphics[width=0.7\textwidth]{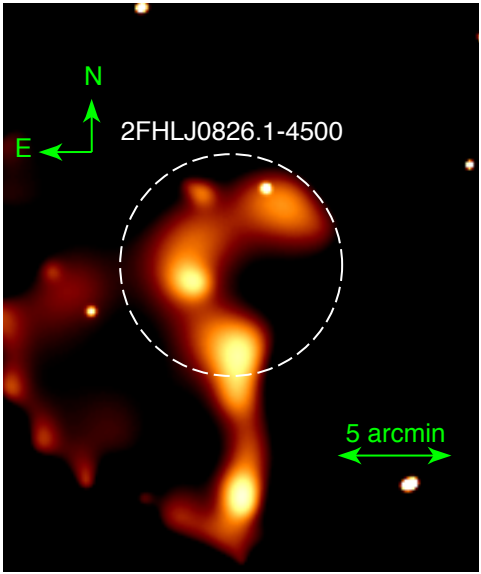}
\end{minipage}
\hspace{-0.8cm}
\begin{minipage}[b]{.5\textwidth}
  \centering
\includegraphics[width=1.2\textwidth]{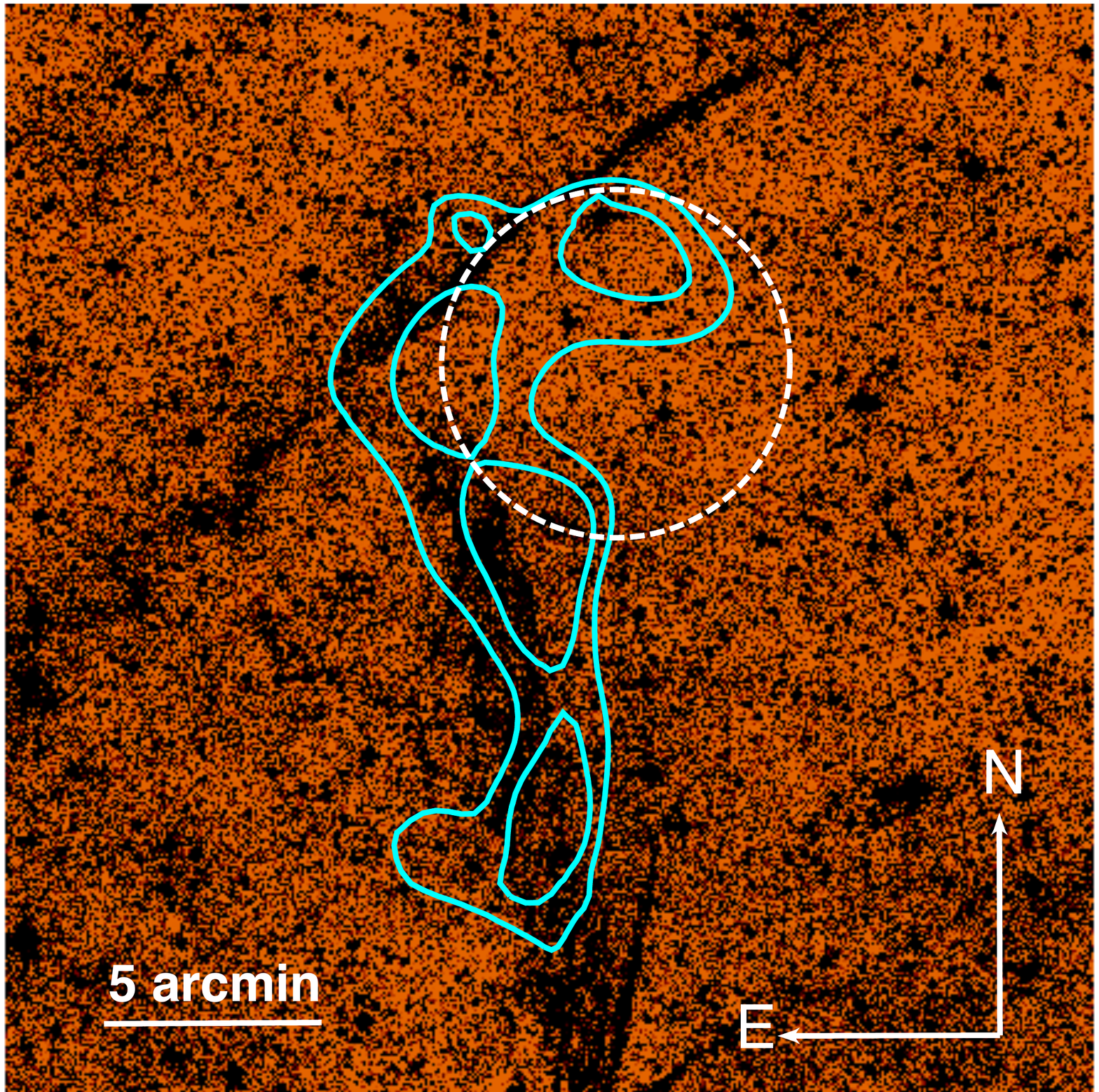}
\end{minipage}
\caption{\textit{Left}: Smoothed, MOS2 0.5--2\,keV image of the region around 2FHL~J0826.1$-$4500. The white dashed circle ($r$=4$^\prime$)  represents the 95\,\% confidence positional uncertainty of 2FHL~J0826.1$-$4500.
\textit{Right}: X-ray emission contours (cyan solid line) overlaid on an H$\alpha$ image of the region of 2FHL~J0826.1$-$4500. The contours are derived from the MOS2 0.5-2\,keV image showed in the left panel and correspond to 1.22$\times$10$^{-2}$ and 1.5$\times$10$^{-2}$ counts.  An optical filament is seen to clearly overlap the X-ray emission. The white dashed circle marks the {\it Fermi}-LAT position.}\label{fig:shock_images}
\end{figure*}

In Figure \ref{fig:shock_images}, left panel, we show the smoothed 0.5--2\,keV image of 2FHL~J0826.1$-$4500, as seen with the MOS2 camera mounted on \xmm. The image was created using the CIAO \citep{fruscione06} tool \texttt{csmooth}, using the fast fourier transforms convolution method and a Gaussian convolution kernel. The minimal signal-to-noise ratio of the signal under the kernel was set to 3.
As illustrated in the X-ray image, in correspondence with the $\gamma$-ray source,  we observe faint, diffuse X-ray emission with extension of roughly 15$^\prime$. 
Furthermore, the X-ray emission is almost spatially coincident with an optical filament visible in a H$\alpha$ image (see Figure \ref{fig:shock_images}, right panel).  
The initial analysis of the \xmm\ diffuse emission reveals it to be very soft, with no significant emission detected  above 2\,keV. 

We perform a spectral fitting in order to find the best model to characterize the observed emission. Usually, the spectral fitting of bright, point-like X-ray sources can be performed subtracting the background emission, since the signal-to-noise ratio of the source is large enough that removing a small fraction of counts from the fitted spectrum does not affect the quality of the analysis. In faint diffuse objects such as 2FHL~J0826.1$-$4500, however, the background subtraction approach can lead to a spectrum with not enough counts to perform a proper spectral analysis. Consequently, the background must be carefully modeled, to then use the best-fit background model as an additional component in the fitting of the total, source plus background, spectrum. In this work, we follow the background modeling approach used in \citet{leccardi}, which takes into account both the instrumental and the astrophysical background. The first is modeled as a combination of quiescent soft protons, cosmic-ray induced continuum, and fluorescence lines; the latter models both the emission from the Galactic Halo and the cosmic X-ray background.

\begin{figure}[htbp]
\centering
\includegraphics[width=.75\linewidth]{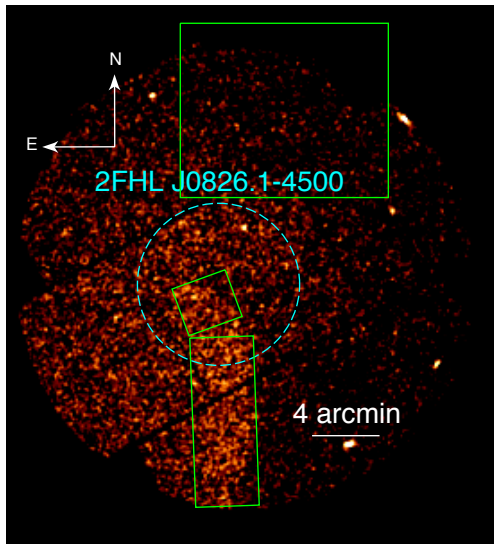}
\caption{Cleaned data from MOS2 CCD with selected regions for source and background emission. The two consecutive green boxes in the lower corner are used for the source and the large green box in the upper right is used for the background.}\label{fig:regions}
\end{figure}

The selected regions for the spectral fitting process are indicated in Figure \ref{fig:regions}. After selecting source and background regions a spectral fitting was then performed with the most recent update of HEASOFT software \citep[v6.19][]{drake2016} with the corresponding calibration files for the \xmm\ telescope\footnote{\label{calib}Current calibration files are accessible at \url{https://heasarc.gsfc.nasa.gov/docs/xmm/xmmhp\_caldb.html}} for MOS1, MOS2, and PN. The resulting spectra were fitted using XSPEC (v12.9.1).

\subsection{Spectral Analysis Results}

We tested two different spectral models, the thermal {\tt mekal} model and a power law, to fit the XMM-\textit{Newton} data of 2FHL~J0826.1$-$4500 and both are reported  in Table~\ref{tab:model}. The Galactic column density has been fixed to $0.026\times10^{22}\,{\rm cm}^{-2}$ \citep{manzali2007} using \texttt{wabs} in XSPEC.
 The  \texttt{mekal} \citep{mewe85,mewe86} model describes the emission spectrum of a hot diffuse gas, assuming as free parameters temperature and metallicity of the gas. In this work, we fix the gas metallicity to the Solar value. The \texttt{apec} model is very similar to \texttt{mekal} and was also considered. With \texttt{apec} we find a result in good agreement with \texttt{mekal} (kT=0.75$^{+0.16}_{-0.20}$\,keV) although with slightly worse best-fit statistics ($\Delta$C-stat=9.7).

The power law model is commonly used to to fit non-thermal spectra and has two free parameters: the photon index and the normalization. More complex power law models, such as \texttt{srcut}, \texttt{cutoffpl} and \texttt{bknpower} are typically found to best characterize synchrotron emission from a non-thermal distribution of electrons: due to the intrinsic faintness of our source, we decide to use the simplest of these models, a pure power law, to minimize the number of free parameters in the fit.

\begin{figure}[!t]
\centering
\includegraphics[width=1.0\linewidth]{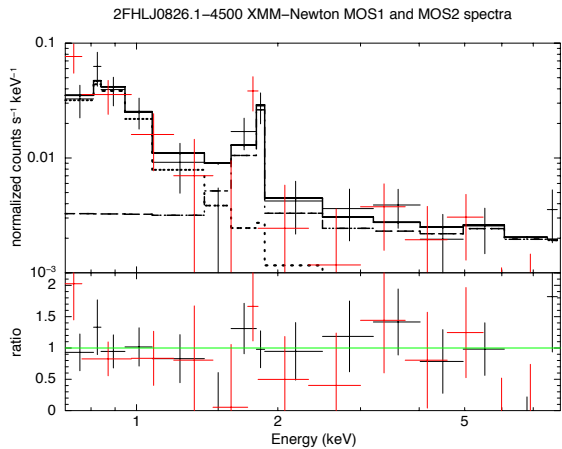}
\caption{\textit{Top}: XMM-\textit{Newton} MOS1 (black) and MOS2 (red) data of 2FHL~J0826.1$-$4500 and the best-fit model obtained using \texttt{mekal}. The best-fit model (solid black line), the instrumental background (dashed black line) and the combination of source and astrophysical background (dotted black line) are plotted. PN data was removed for clarity.}\label{fig:mekal}
\end{figure}

As shown in Table \ref{tab:model}, a thermal emission scenario (Cstat/d.o.f.=454.63/449) is statistically preferred to a non-thermal one (Cstat/d.o.f.=470.60/451). 
The best-fit temperature is $kT=0.60^{+0.11}_{-0.60}$\,{\rm keV} with an upper limit of $kT<0.72\,keV$. We show the  best fit model, as well as our MOS1 and MOS2 data, in Figure \ref{fig:mekal}. PN data has been removed for clarity.
The observed 0.5--2\,keV source flux, without taking into account the different background contributions, is $<$1.9$\times$10$^{-13}$\,erg s$^{-1}$ cm$^{-2}$.

\begingroup
\renewcommand*{\arraystretch}{1.2}
\begin{table*}
\centering
\scalebox{1.}{
\begin{tabular}{cccccccc}
\hline
\hline
\ Spectral Model & $\chi^2$ & C-Stat & d.o.f.$^a$ & Reduced $\chi^2$ & $kT$(keV) & Photon Index \
\\
\hline
mekal & 458.54 & 454.63 & 449 & 1.02 & 0.60$_{-0.60}^{+0.11}$ & - \\
power law & 472.60 & 470.60 & 451 & 1.05 & - & 4.9$_{-1.2}^{+1.8}$ \\
\hline
\hline
\end{tabular}}\caption{Summary of the best-fit parameters and the associated statistics for both spectral models used in our analysis. Because of the low qualiy of data, the $kT$ value only generates an upper limit of 0.72\,keV. {\footnotesize \textit{$^a$ degrees of freedom}}}
\label{tab:model}
\end{table*}
\endgroup

\section{Multi-wavelength Information}\label{sec:multi}
\subsection{Soft X-rays}
\begin{figure*}[]
\centering
\includegraphics[width=0.9\linewidth]{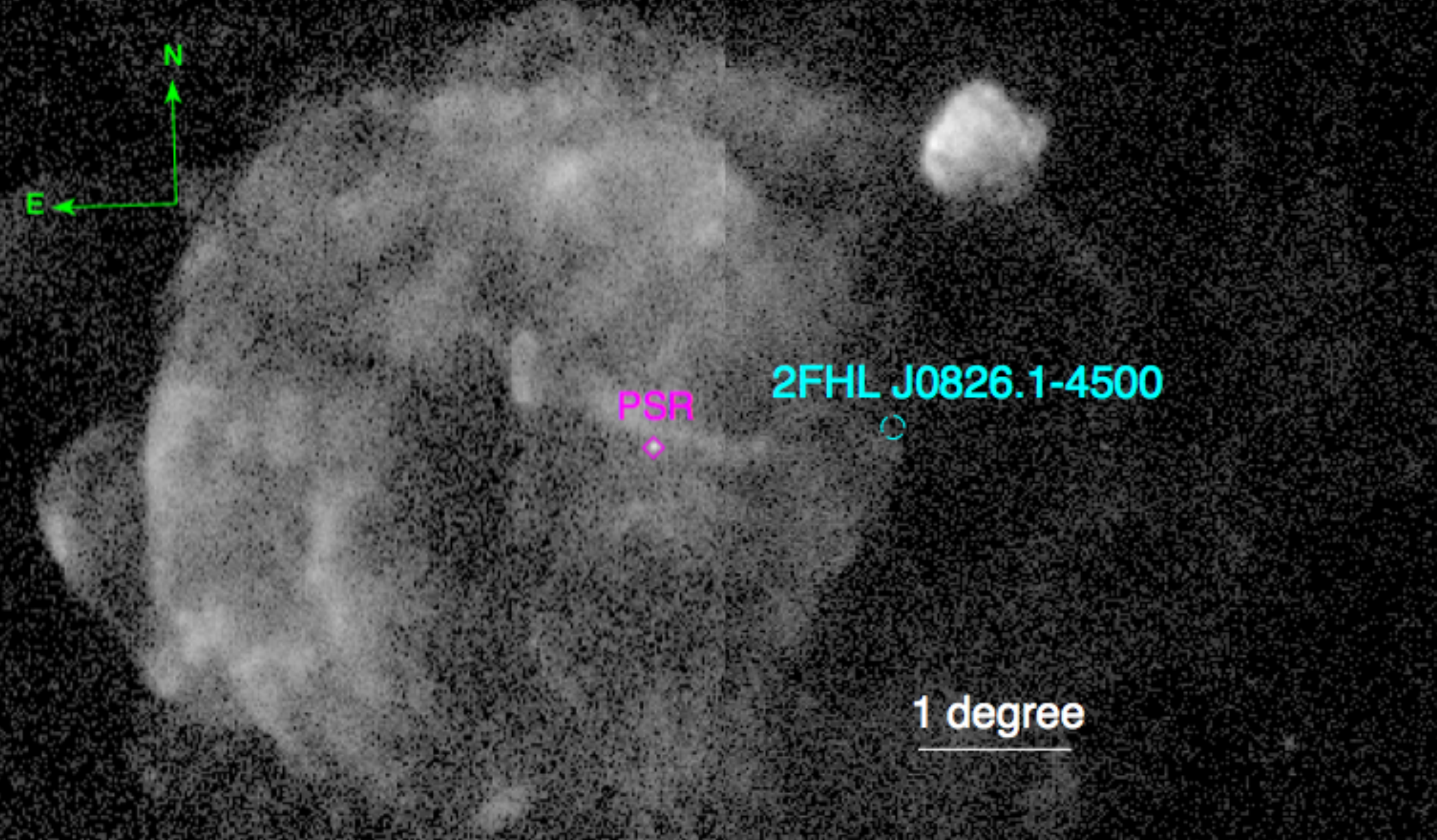}
\caption{0.5--2.4\,keV ROSAT image of the Vela SNR. The cyan circle labels the $\gamma$-ray location of 2FHL~J0826.1$-$4500 (with uncertainty of $r$=4$^\prime$ at 95\,\% confidence), while the magenta diamond outlines the position of the Vela pulsar. 2FHL~J0826.1$-$4500 lies on the western edge of a prominent X-ray shell, just before a large cavity. The bright source in the top right corner is Puppis A.}\label{fig:rosat}
\end{figure*}

The Vela SNR is one of the brightest, largest soft (0.5-2\,keV) X-ray sources. The ROSAT X-ray telescope \citep{rosat1999} has mapped the Vela region in the 0.5--2.4\,keV band, showing extended emission over a $\sim$8$\degree$ region that encompasses the Vela pulsar and roughly outlines the SNR shell. 2FHL~J0826.1$-$4500 lies just on an inner boundary of X-ray emission that delineates the SNR shell where, to the west, a cavity of little X-ray emission is present (Figure \ref{fig:rosat}).
\citet{bach2000} performed a spectral analysis on 3 distinct regions of the SNR and found that the X-ray emission is best fit by a thermal emission model with two Raymond-Smith plasma components with different temperatures. The best fit temperatures are 0.12, 0.17, and 0.18\,keV for $T_{1}$ and 0.76, 1.06, and 0.82\,keV for $T_{2}$ in one bright region in the north and two fainter regions in the north-east and south, respectively. These results are in good agreement with those obtained by our XMM-\textit{Newton} observation and reported in the previous section.

\subsection{$\gamma$-rays}
 At $\geq$50\,GeV, 2FHL~J0826.1$-$4500 is a relatively faint source detected by the \fermi{} with a test statistics of $\sim$27 (corresponding to $\sim$4.5\,$\sigma$) and  only $\sim$5 photons \citep{ackermann2015}. In the 2FHL catalog, no evidence of extended emission at $>$50\,GeV is reported for 2FHL~J0826.1$-$4500, although the significance of this result is limited by the small number of counts detected by the LAT.



\citet{hess2012} found TeV emission up to 10\,pc ($\sim$1.2$\degree$ from pulsar position) from the pulsar on the opposite side of the SNR with respect to 2FHL~J0826.1$-$4500.  At 1\,TeV, the angular resolution for H.E.S.S. is $\sim$6$^{\prime}$. The shock is approximately 17$^\prime$ in length in the X-rays, and it is observed by LAT up to $\sim$400\,GeV (see Figure \ref{fig:sed_gamma_map}, left panel), so it is intriguing that \citet{hess2012} did not detect any significant emission in the TeV regime at the position of 2FHL~J0826.1$-$4500. The lack of HESS detection implies either that the spectrum of the source does not extend beyond 1\,TeV or that the source is variable. It is unlikely the detection of source variability in this case will be possible with the \fermi{} because of the low photon counts of 2FHL~J0826.1-4500. The recently published HESS Galactic Plane Survey  \citep{hess2018} allows us to derive a 5\,$\sigma$ upper limit on the emission of 2FHL~J0826.1$-$4500 (assuming it is a point-like source) at $>$1\,TeV of  2.2$\times 10^{-13}$\,erg s$^{-1}$ cm$^{-2}$.

\subsection{Radio}
\begin{figure*}
\begin{minipage}[b]{.5\textwidth}
  \centering
\includegraphics[width=0.9\textwidth]{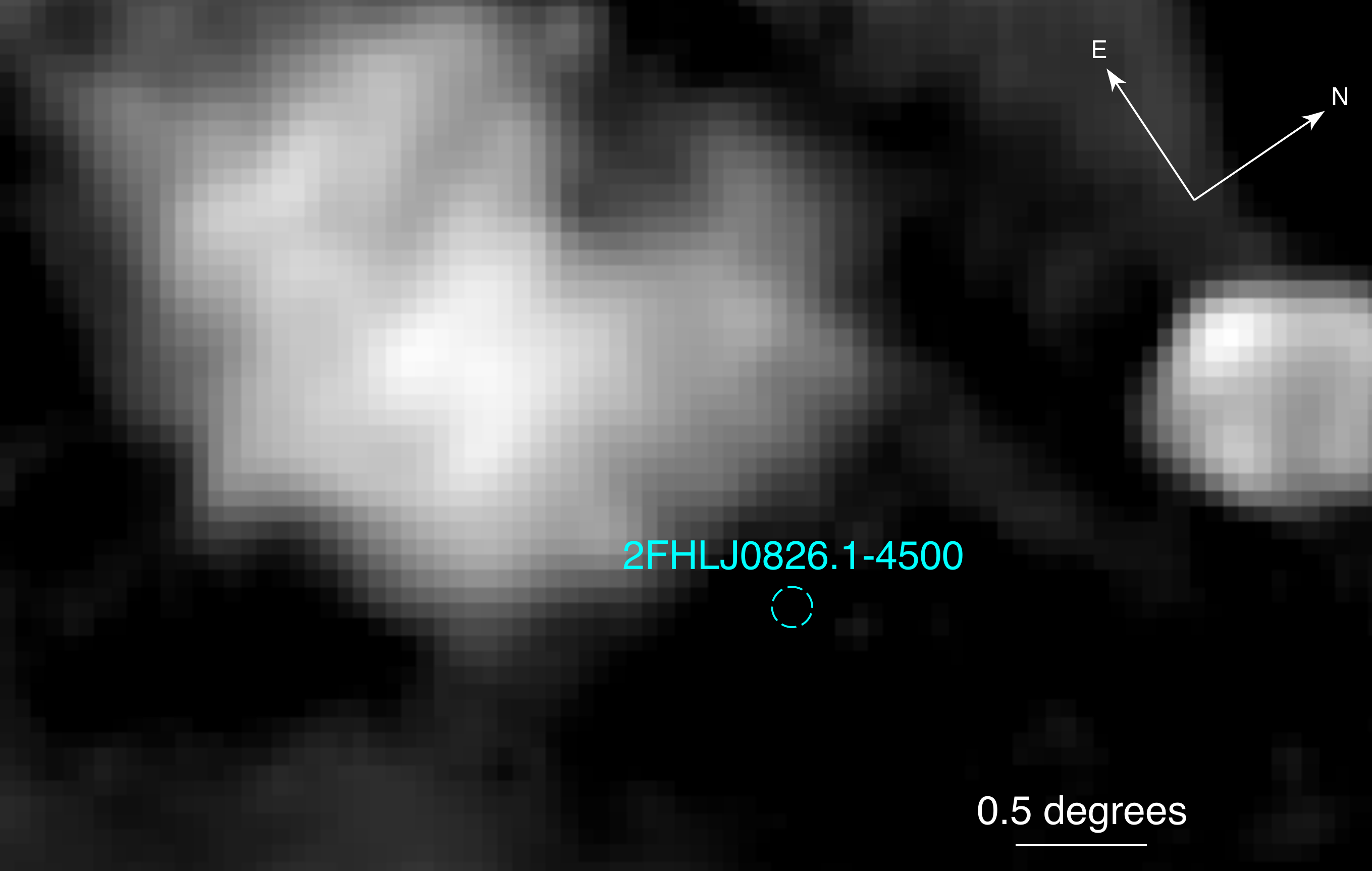}
\end{minipage}
\hspace{-0.4cm}
\begin{minipage}[b]{.5\textwidth}
  \centering
\includegraphics[width=0.9\textwidth]{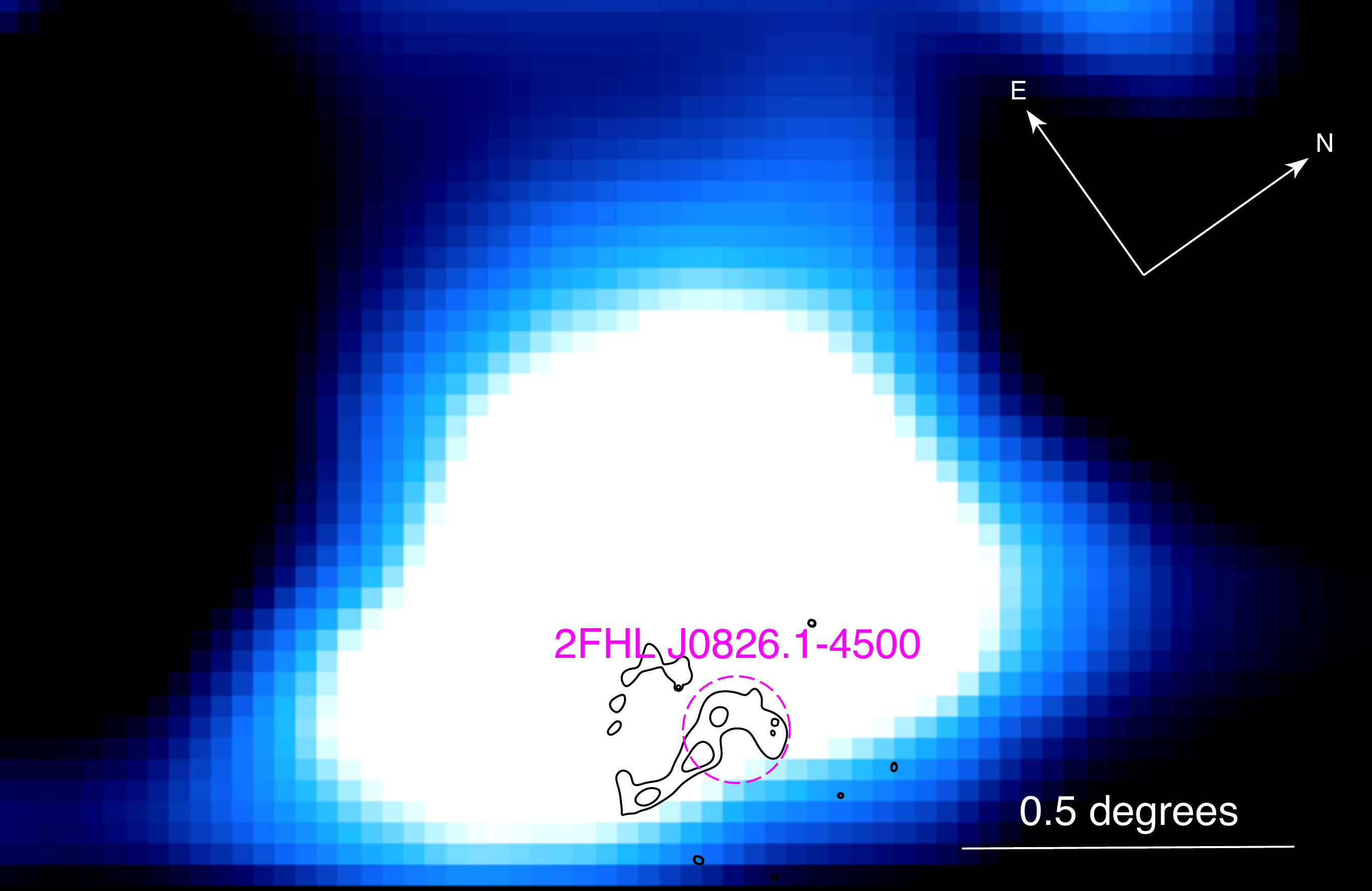}
\end{minipage}
\caption{\textit{Left}: 2.4GHz radio emission map from \citet{duncan1996}$^{\ref{duncan}}$ with location of 2FHL~J0826.1$-$4500 seen in a cavity with no radio emission detected just below the Southwestern corner of the PWN, Vela-X.
\textit{Right}: HI 21cm radio map integrated between 29.7 and 35.3\,km s$^{-1}$ indicating the location of 2FHL~J0826.1$-$4500 with respect to the HI cloud with black contours for reference of shock structure and location \citep[see][for a review]{dubner1998}.
}\label{fig:duncan_images}
\end{figure*}

The Vela region is rich with radio emission, predominantly coming from the Vela-X PWN, with lower surface brightness emission roughly outlining the SNR X-ray shell. \citet{duncan1996} investigated the Vela radio emission at 2.4\,GHz: when we positionally match their data with our XMM-\textit{Newton} image we find that the X-ray emission lies on the outskirts of the diffuse radio emission. As can be seen in Figure \ref{fig:duncan_images} (right panel), 2FHL~J0826.1$-$4500 lies in a region lacking radio emission\footnote{\label{duncan}The radio maps reported in \citet{duncan1996} are available at \url{http://www.atnf.csiro.au/research/surveys/2.4Gh\_Southern/data.html}}. This evidence is confirmed throughout different radio observations at different frequencies. For example, observations of the $^{12}$CO emission at 115GHz revealed the same behavior of the Vela region \citep{mori2001}. 

The apparent lack of radio emission from the SNR at the position of source 2FHL~J0826.1$-$45.00 and further west could be related to the forward shock having broken out into a lower density ISM region. This is also consistent with observations of the pulsar position and proper motion, since the pulsar is known to be in the Northern part of the radio emission of the PWN. The offset is likely caused by the expansion of the SNR into the ISM and how this interacts with the PWN. It is likely that a non-uniform density in the ambient ISM is the cause of the asymmetric appearance \citep{blondin_2001,slane_2018}, and is responsible for an early return of the reverse shock of the SNR from the direction where the shock front encounters a denser medium \citep[i.e., in this case, the structure Northern edge, see][]{slane_2018}. The atomic hydrogen density for the Northern edge of the SNR, n=1--2\,cm$^{-3}$, is indeed greater than the one estimated for the Southern edge, n$\sim$0.1\, cm$^{-3}$, supporting the scenario  just described \citep{dubner1998, hess2012, jager2008}.

\citet{dubner1998} mapped 21cm HI clouds in the Vela region reporting negative radial velocities (-21 km s$^{-1}$ to -9 km s$^{-1}$). Later \citet{bach2000} connected the high $N_{H}$ column densities in the Southern region with the HI clouds observed at 21 cm, concluding that these clouds must be moving towards the observer and are likely being accelerated by the SNR shock wave. \citet{dubner1998} interpreted the HI presence as a bubble that the Vela SNR is expanding into, with a higher interaction in the North and East directions between the HI clouds and forward shock. \citet{bach2000} have speculated that the interaction is weaker with HI in the South and West directions, allowing the shock wave to keep expanding inside the bubble. This may support the idea that, at the position of 2FHL~J0826.1$-$4500, the ROSAT X-ray boundary may be confined by the interaction with a small HI cloud. \citet{bach2000} also suggests that the high $N_{H}$ measured implies that a region of cold interstellar gas lies behind the Southern boundary of the SNR, and that it is likely for the HI clouds to be in front of the remnant in the West, which may explain the apparent cavity that exists West of the XMM-\textit{Newton} X-ray source.

Furthermore, a small HI cloud has been identified in \citet{dubner1998}, widely overlapping with the H$\alpha$ optical filament associated with the observed \xmm\ emission (see Figure \ref{fig:duncan_images}, left panel). The morphology of the filament correlates well with the location and size of the HI cloud in this region, suggesting that we may be observing a shock-cloud interaction, with the shock being visible in the the optical band and in X-rays. Notably, optical emission in SNRs is usually associated to bright X-ray boundary regions, which is indeed what is observed on the ROSAT map. This may suggest a density enhancement in the Western region of the Vela SNR, providing further evidence for a forward shock scenario. 

In conclusion, the combined $\gamma$-ray, X-ray, optical and radio information depicts a scenario of interaction between a forward shock, linked to the SNR, and a HI cloud. The following section will discuss this scenario and its implications. 

\section{Discussion}\label{sec:discuss}
\subsection{Shock-Cloud Interactions}
\citet{castro2015} examined the X-ray and $\gamma$-ray signatures of SNR interactions with molecular clouds suggesting several ways to confirm if a shock-cloud interaction is indeed occurring. Most SNRs are located within high density molecular cloud regions making shock-cloud interactions a likely occurrence with SNRs as they expand into the ISM. Morphology such as arcs, curvature, and asymmetric appearance of the SNR provide suggestive clues of an interaction with surrounding medium especially if a correlation with the shape of a nearby cloud can be established \citep{castro2015}. 
The shape of 2FHL~J0826.1$-$4500 has a compelling overlap with the shape of the HI cloud as can be seen in Figure \ref{fig:duncan_images}, left panel. 

Multi-wavelength studies like the one here are another ideal way to confirm if a shock-cloud interaction is in fact happening. CO emission presence, for instance, is an efficient tool to map the distributions and movements of dense clouds, and is commonly used to confirm the physical interaction of a SNR with a molecular cloud \citep{castro2015}. 

Gathering evidence for radiative processes at the shock site will provide strong indicators for the speed of shock front. Mapping the presence of OIII, NII, or SiII optical line emission can be strong evidence for a radiative shock that is propagating into an inhomogeneous ISM \citep{castro2015}. More specifically, if OIII, NII, or SiII are present the shock would be slower than the blast wave velocity as it comes into contact with a dense cloud. If the velocity of the shock is still considerably fast then the shock may be just starting to interact with dense material. Maser emission at 1720MHz would also be direct evidence of excitation occurring at a site where a shock and cloud are interacting.

\citet{winkler2014} provides substantial evidence linking shock-cloud interactions to the presence of optical filaments that closely overlap X-ray emission in the SNR SN~1006. Analogously, the close coincidence between the H$\alpha$ emission morphology and that of the X-ray emission in 2FHL~J0826.1$-$4500 suggests that part of the SNR shock is interacting with a region of partially neutral material in this area.

\begin{figure*}
\centering
\includegraphics[width=0.9\linewidth]{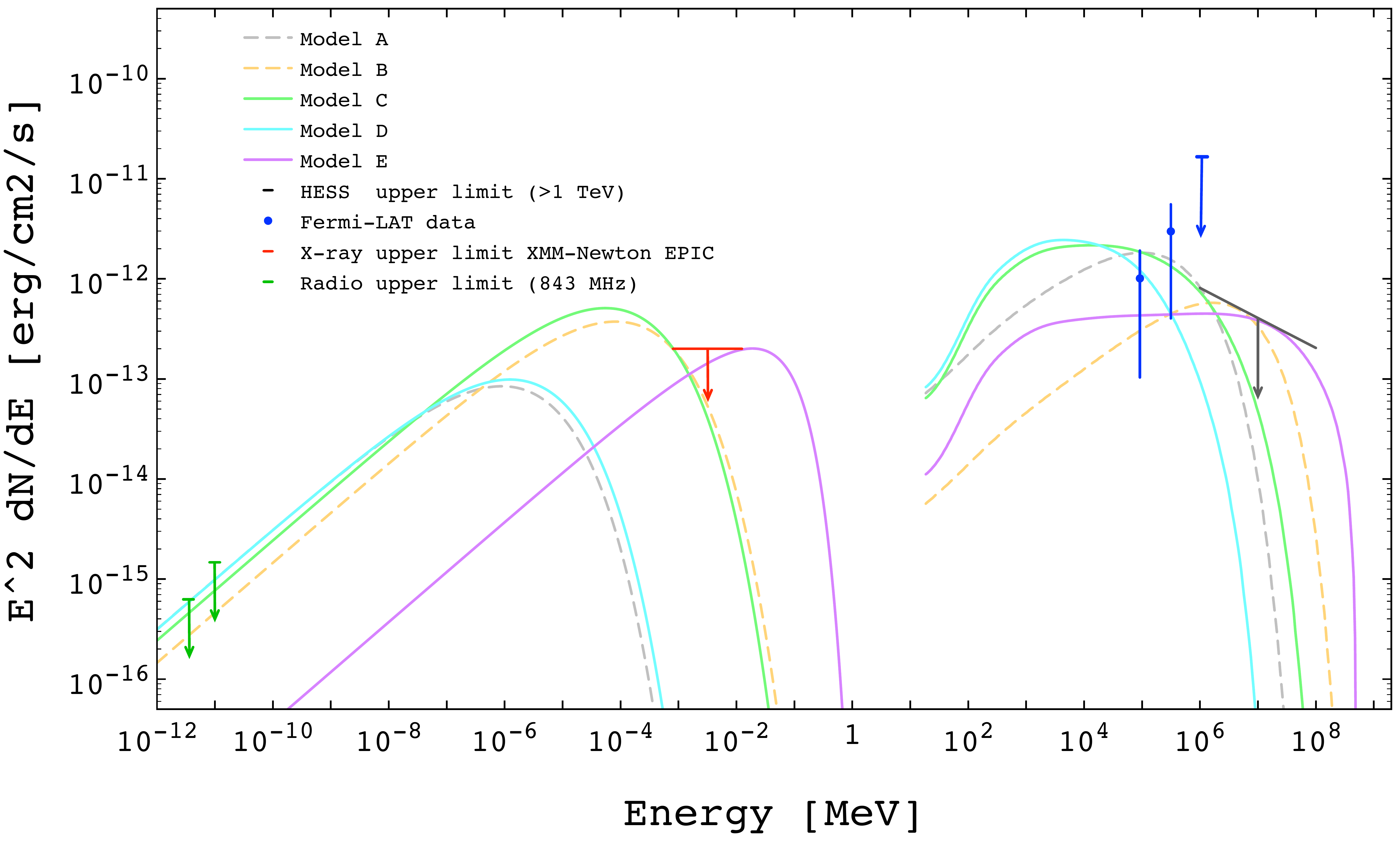}
\caption{Spectral energy distribution (SED) for various scenarios constrained to upper limits of available data across the electromagnetic spectrum. Models A (gray dashed line) and B (yellow dashed line) demonstrate resultant $\gamma$-ray spectrum of radiation from relativistic electrons. Models C (solid green), D (solid cyan), and E (solid purple) demonstrate resultant spectrum of radiation from a hadronic population.}\label{fig:spec}
\end{figure*}

\subsection{Efficient Particle Acceleration}

SNRs are widely thought to accelerate a significant fraction of Galactic CRs through diffusive shock acceleration in their high-velocity blastwaves. The $\gamma$-ray emission in the MeV-GeV band from regions with relatively high ambient density is expected to be hadronic in origin \citep[see][Figure 6, for an example]{castro_2013b}, and hence evidence of CR hadron acceleration at these shocks. Proton-proton collisions between shock accelerated CR ions and ambient protons are enhanced in high density regions such as an interaction between a SNR forward shock and an HI cloud. The energetics of 2FHL~J0826.1-4500 make this site a promising source for efficient particle acceleration to CR energies. The necessary system parameters for both leptonic and hadronic emission scenarios are investigated in \S 4.3.

Regarding the origin of the underlying particle population, it is difficult to distinguish whether the emission from this shock-cloud region originates from freshly shock-accelerated CRs, or instead from pre-existing CRs re-accelerated through the high compression ratio characteristic of radiative shocks. 
There are two observational properties that help differentiate between these two scenarios. First, one would expect the $\gamma$-ray spectrum from re-accelerated field CRs to have a lower energy cut-off value relative to that of emission from diffusive shock accelerated CRs \citep{ackermann2013,lee2015}. 
Fast shocks typically reveal a potential for fresh CR acceleration and the speed of the shock can be indicated by the $\gamma$-ray spectrum cutting off at energies greater than 20\,GeV (e.g. SNR Cas A; see \citet{magic2017}, RX~J1713.7-3946; see \citet{hessrx}, RX~J0852.0-4622; see \citet{aha2007}, \citet{bykov2018}, and \citet{abdalla2018}). For slow shocks this cut off occurs at energies of $E<$10-20\,GeV, indicating that the site is most likely re-accelerating pre-existing CR protons (e.g. 1FGL~J1801.3-2322c; see \citet{abdo2010}, SNR W44; see \citet{w442011} and \citet{w442016}, G349.7$+$0.2 and CTB~37A; see \citet{castro2010}). This effect is compounded by Alfven-wave damping as the radiative shock propagates through significantly neutral environments, which introduces a break in the particle momentum spectrum, further softening it \citep[see][and references therein]{lee2015}.
We investigate the possible values of the cut-off energy of the particle spectrum in the site of 2FHL~J0826.1-4500 in \S 4.3.

Additionally, the X-ray band might also provide a clue since SNRs interacting with molecular clouds where the shock has become radiative and that are bright in the GeV band, such as W44 and IC443, are characterized by a center-filled X-ray morphology, rather than a shell-like one \citep{castro2015}. Since thermal X-ray emission is associated with bright optical filaments in the region of 2FHL~J0826.1-4500, it appears as though the shock is still fast enough to heat the surrounding medium to X-ray emitting temperatures, and hence a significant part of the shock is likely non-radiative. This suggests that the CRs in this region have been produced through diffusive shock acceleration. However, a deeper analysis in VHE and optical is required in order to confirm this.

\subsection{Modeling Spectral Energy Distribution}

The multi-wavelength information available can be combined to build a picture of the broadband spectral characteristics of the region. Assuming the GeV $\gamma$-ray emission in the direction of \fhl\ is indeed the result of radiation from a relativistic particle population accelerated at a region of the Vela SNR shock, it is possible to model the broadband emission from the shock-accelerated non-thermally distributed electrons and protons and hence derive constraints on the physical parameters of the shock. The data of the region are shown in Figure~\ref{fig:spec}, where the 843 MHz and 2.4 GHz radio upper limits are derived from \citet{murphy_2007} and \citet{duncan1996} respectively, the X-ray upper limit is obtained from the \xmm\ observations as described in \S 2, and the TeV $\gamma$-ray upper limit is from the H.E.S.S. Galactic Plane Survey \citep{hess2018}.

We assume the distribution of the accelerated particles in momentum to be $dN_{i}/{dp} = a_{i} \,p^{-\alpha_{i}} \exp\left(-{p}/{p_{0\,i}}\right)$.
Here, subindex $i$ represents the particle type (proton or electron), and $\alpha_{i}$ and $p_{0\,i}$ are the spectral index and the exponential cutoff momentum of the distributions. The coefficients for the particle distributions, $a_{p}$ and $a_{e}$, are set using the total energy in relativistic particles and the electron to proton ratio as input parameters, together with the spectral shape of the distributions. The spectral indices of electron and proton distributions are assumed to be equal since analytic and semi-analytic models of particle acceleration at shocks suggest this is the case \citep{reynolds_2008}. For the non-thermal radiation from these particle distributions we have used $\pi^0$-decay emission from \citet{kamae_2006,mori_2009}, synchrotron and inverse Compton (IC) emission from \citet[][and references therein]{baring_1999}, and non-thermal bremsstrahlung emission from \citet{bykov_2000}. For more details on the model for the particle distribution and their simulated emission see \citet{castro_2013}. 

\begin{table}
\centering

\begin{tabular}{ccccccccc}
\hline
\hline
\noalign{\vskip 1mm} 

\multirow{2}{*}{}&\multirow{2}{*}{}&$p_{\text{0}}$&$B_{2,max}$&$n_{\text{H}}\times E_{\text{CR,p}}$&$E_{\text{CR,e}}$\\

&&{(TeV/c)}&($\mu$G)&{($10^{48}$ erg/cm$^{3}$)}&{($10^{44}$ erg)}\\
\noalign{\vskip 1mm} 
\hline
\noalign{\vskip 1mm} 
\multirow{2}{*}{\it Leptonic}&{\it A}  &  5   & 3  & --   & 9 \\ 
&{\it B}  &  30    & 9  & --   & 1 \\ 
\noalign{\vskip 1mm} 
\hline
\noalign{\vskip 1mm} 

\multirow{3}{*}{\it Hadronic}&{\it C}  &  10    & 50   & 6 & -- \\ 
&{\it D}  &  1   & 50   & 6 & -- \\ 
&{\it E}  &  600   & 10   & 1  & -- \\ 



\noalign{\vskip 1mm} 
\hline
\noalign{\vskip 1mm} 

\end{tabular}
\caption{Input Model Parameters.
}\label{tab:models}
\end{table}



We use the model outlined above to establish the approximate ranges of some of the physical parameters that would result in emission that fits the \fermi\ data, as well as complying with the upper-limits at other wavelengths. The common parameters for all models considered are a relativistic electron to proton ratio of $k_{ep}=0.01$ (determined in observations of CR abundances on Earth), and a shock compression ratio of 4.
Both of these standard assumptions are discussed in more depth in \citet{castro_2013}. It is possible that the proximity of the Vela pulsar and pulsar wind nebula, both of which are expected to be significant electron accelerators, might increase the ratio of relativistic electrons to protons present in the region of interest. A larger number of CR electrons would relax the constraints on the magnetic field placed on both leptonic and hadronic scenarios considered here. However, this effect is presumably not very significant since the $\gamma$-ray source appears to be coincident with the optical and X-ray emitting shock, and hence presumably the CRs responsible for the emission are accelerated in this same region. Additionally, we adopt a distance of $d=0.29$ kpc (the distance of the Vela SNR, as discussed above), and fix the spectral index of the relativistic proton and electron distributions in momentum to be $\alpha_i=4$. This last assumption is adopted given that neither the radio nor the $\gamma$-ray observations allow for a tightly constrained spectral index of the emission spectrum and hence, we default to the canonical spectral index expected from diffusive shock acceleration \citep[see ][and references therein]{reynolds_2008}. The input parameters for each model considered are included in Table \ref{tab:model} and the resulting broadband spectral distributions are shown in Figure \ref{fig:spec}. 

In models A and B, the \fermi\ spectrum is the result of radiation from relativistic electrons (leptonic channel), and in C, D and E, it originates from protons and heavier ions colliding with ambient hadrons and resulting in $\gamma$-ray emission from pion decay (hadronic models). Models A and B represent the highest and lowest values, respectively, for the energy CR electrons, $E_{\text{CR,e}}$, allowed by the \fermi\ data. The post-shock magnetic field, $B_2$, in each of these two cases, is set to the maximum value possible without the synchrotron emission exceeding the radio and X-ray limits, and the cut-off of the particle momentum distribution, $p_{\text{0}}$, is limited by the upper-limit on the TeV $\gamma$-ray emission from H.E.S.S. From these cases we estimate that if the GeV emission is the result of leptonic processes, the total CR electron energy in this region must be roughly 1--9 $\times10^{44}$ erg, and the cut-off momentum of the particle distribution must be approximately 5--30 TeV$/$c. Additionally, the maximum values of the magnetic field strength seem to be between 3 and 9 $\mu$G, which are low given that one would expect values larger than 12 $\mu$G at sites of efficient particle acceleration \citep[see][and references therein]{castro_2013b}.

Pion decay emission is proportional to both the amount of energy in CR protons in the region, and the post-shock density \citep{kamae_2006}. Hadronic models C and D both result from the maximum values of the product $n_{\text{H}}\times E_{\text{CR,p}}$ allowed by the 2FHL data, and model E represents the case with the minimum $n_{\text{H}}\times E_{\text{CR,p}}$ possible. In models C the cut-off in proton momentum is at the highest value allowed by the TeV upper limit, for the maximum $n_{\text{H}}\times E_{\text{CR,p}}$ cases. In turn, model D is the result of the minimum $p_{\text{0}}$ that still fits the GeV data. From these hadronic models we estimate the range of $n_{\text{H}}\times E_{\text{CR,p}}$ to be $\sim1-6\times10^{48}$ erg/cm$^{3}$, the cut-off of the momentum distribution to be approximately $1-600$ TeV/c, and the maximum post-shock magnetic field strength $B_2\sim50$.










\section{Conclusions}\label{sec:conclude}

The discovery and investigation of a likely shock-cloud interaction taking place on the western edge of the Vela SNR is presented. Multi-wavelength data suggests the forward shock of the SNR is interacting with a small HI cloud as a likely scenario. The data presented for 2FHL~J0826.1$-$45.00 points towards the possibility of a site for CR acceleration. A broadband spectral fitting is reported for several particle populations that can explain the emission observed including leptonic and hadronic scenarios. If the hadronic models prove to be most realistic in characterizing 2FHL~J0826.1$-$45.00 across available data, 2FHL~J0826.1$-$45.00 is not only a site of efficient particle acceleration but also poses as a candidate for fresh CR acceleration.

Future work to further characterize 2FHL~J0826.1$-$4500 includes studying the physical conditions (velocity, direction, elemental composition) of the shock front which will be valuable in better characterization of the region in order to understand its kinematics. These data will provide clues to whether the western edge of the Vela SNR may be a site generating fresh CRs or is instead a re-acceleration mechanism.


\bibliographystyle{aa}
\bibliography{apj_paper_layout}

\end{document}